# Integrating Privacy Enhancing Techniques into Blockchains Using Sidechains

Reza M. Parizi, Sajad Homayoun, Abbas Yazdinejad, Ali Dehghantanha, Kim-Kwang Raymond Choo

Department of Software Engineering and Game Development, Kennesaw State University, GA, USA
rparizi1@kennesaw.edu
Department of Computer Engineering and Information Technology, Shiraz University of Technology, Shiraz, Iran
s.homayoun@sutech.ac.ir
Faculty of Computer Engineering, University of Isfahan, Iran
abbasyazdinejad@yahoo.com
School of Computer Science, University of Guelph, Ontario, Canada
adehghan@uoguelph.ca
Department of Information Systems and Cyber Security, The University of Texas at San Antonio, TX, USA
raymond.choo@fulbrightmail.org

*Abstract*—Blockchains are turning into decentralized computing platforms and are getting worldwide recognition for their unique advantages. There is an emerging trend beyond payments that blockchains could enable a new breed of decentralized applications, and serve as the foundation for Internet's security infrastructure. The immutable nature of the blockchain makes it a winner on security and transparency; it is nearly inconceivable for ledgers to be altered in a way not instantly clear to every single user involved. However, most blockchains fall short in privacy aspects, particularly in data protection. Garlic Routing and Onion Routing are two of major Privacy Enhancing Techniques (PETs) which are popular for anonymization and security. Garlic Routing is a methodology using by I2P Anonymous Network to hide the identity of sender and receiver of data packets by bundling multiple messages into a layered encryption structure. The Onion Routing attempts to provide low-latency Internet-based connections that resist traffic analysis, de-anonymization attack, eavesdropping, and other attacks both by outsiders (e.g. Internet routers) and insiders (Onion Routing servers themselves). As there are a few controversies over the rate of resistance of these two techniques to privacy attacks, we propose a PET-Enabled Sidechain (PETES) as a new privacy enhancing technique by integrating Garlic Routing and Onion Routing into a Garlic Onion Routing (GOR) framework suitable to the structure of blockchains. The preliminary proposed GOR aims to improve the privacy of transactions in blockchains via PETES structure.

## I. INTRODUCTION

Blockchain is based on distributed ledger technology, which securely records information across a peer-to-peer network. In spite of the fact that it was initially made for exchanging Bitcoin, blockchain's potential reaches far beyond cryptocurrency. The presentation of Bitcoin protocol by Satoshi Nakamoto [1] in 2008 has stamped the starting of this new period of decentralization in computer program frameworks worldwide.

The advantages of blockchains (e.g., high security and transparency) do however come at a price [2]. There is a privacy concern that blockchain technology is incompatible with Data Protection Regulation and ACT, which data must be kept private at the user's behest [3]. In the current blockchain environments, the issue is not only that data is permanently stored on a ledger, never to be erased, but that by nature it exists on a blockchain which is irreversibly shared with the entire network. Although the pseudonym nature of blockchains is deemed to protect the users' identity [4], reputation and privacy is a very challenging issue because it is still possible to trace back transactions to a specific identity through blockchain analytics techniques. As blockchain technology generally requires the storage and exchange of data on a large scale, it creates new concerns with the "Reputation and Privacy" of its users, and it also increases privacy compliance concerns for organizations adopting this new technology. Furthermore, the new applications of blockchain could even increase the privacy concerns of reputation and privacy in the near future.

Protecting the privacy of financial transactions has long been a goal of the cryptography community [5]. Many cryptocurrencies, such as Bitcoin, do not provide true anonymity: transactions include pseudonymous addresses, meaning a user's transactions can often be effortlessly connected. As a trivial solution, users may follow creating new addresses, but this does not make transactions anonymous as all transfers are globally obvious within the blockchain and will be available forever. Several recent studies have considered ways to connect a user's addresses to each other and to an outside identity [6-9].

This paper introduces PET-Enabled Sidechain (PETES) as a new privacy enhancing technique for improving the privacy of cryptocurrency transactions by integrating the proposed PET functionality into a sidechain. Our PETES takes the advantage of both Onion Routing [10] and Garlic Routing [11] into a Garlic Onion Routing (GOR) technique.

The remainder of this paper is organized as follows. Section II reviews related research, while Section III gives the background knowledge. We describe our approach for improving privacy of transactions in Section IV. Finally, Section V concludes the paper.

## II. RELATED WORK

There are many techniques used for anonymization of transactions and obfuscating the identity of an individual using blockchain technology. Coin-mixing [12] as an anonymization technique takes the advantage of third-party to remove the connecting link between an address sending a form of payment and the address it is intended for. Coin-mixing is useful in the context of Bitcoin or similar services that have digital identities present (e.g. transactions involving digital assets). Within the case of Bitcoin, coin-mixing is imperative because it will degrade the possibility of anybody to see where transactions are coming from and being sent as well, that can be connected to a particular client account(s) [13]. A few cryptocurrencies, such as Zcash [14], are now supporting anonymous payment scheme on top of blockchains using Transaction Remote Release (TRR) [15] that follows a layer by layer encryption scheme of Tor [16] in which the sender encrypts the new transaction utilizing the public keys of a few TRR clients (Bitcoin clients utilizing TRR protocol). A TRR node decrypts the received data using its private key and sends the data to the next TRR node. The last node publishes the transaction to the network [6, 17].

The authors in [18] construct a full-fledged ledger-based digital currency with strong privacy guarantees, named Zerocash. Their results leverage recent advances in zero-knowledge Succinct Non-interactive Arguments of Knowledge (zk-SNARKs). They formulate and construct decentralized anonymous payment schemes (DAP schemes). A DAP scheme enables users to directly pay each other privately in Bitcoin platforms: the corresponding transaction hides the payment's origin, destination, and transferred amount. Similarly, the work in [19] explores the role of privacy-enhancing overlays in Bitcoin. To inquire the effectiveness of diverse solutions, the authors propose a formal definitional framework for virtual currencies and put forth a new notion of anonymity, taint resistance that they can satisfy.

Monero [20] is a privacy-centric cryptocurrency that allows users to obscure their transactions by including chaff coins, called "mixins," along with the actual coins they spend. The authors propose and evaluate two countermeasures that can improve the privacy of future transactions.

R. Mercer [21] introduces a unique ring signature (URS) scheme that works with existing blockchain systems for privacy enchantment. He implemented the scheme using secp256k1 and compared its efficiency with other commonly suggested approaches to privacy on the blockchain. The authors in [22] have offered an efficient NIZK Scheme for Privacy- Preserving Transactions over Account-Model Blockchain that utilizes a homomorphic public key encryption scheme and construct a highly efficient non-interactive zero knowledge (NIZK) argument based upon the encryption scheme to ensure the validity of the transactions.

III. BACKGROUND

This section provides a brief background knowledge on the underlying techniques used in our approach.

*A. Onion Routing*

The Onion Routing attempts to provide low-latency Internet-based connections that resist traffic analysis, de-anonymization attack, eavesdropping, and other attacks both by outsiders (e.g. Internet routers) and insiders (Onion Routing servers themselves) [10]. Onion Routing relies on using Public Key Cryptography [23], which allows it to encrypt layers of onions such that only intended recipients of each layer can decrypt it with their private keys. Onion routing is implemented by encryption in the application layer of a communication protocol stack, nested like the layers of an onion. Tor [10] is currently the most advanced implementation of Onion Routing and is an anonymous network overlaid on the public internet that allows its users to anonymously access the internet, and to use internal Tor websites that reside only within the Tor network. Fig. 1 shows that Onion Routing encapsulates data to pass from the predefined path of routers from source to destination.

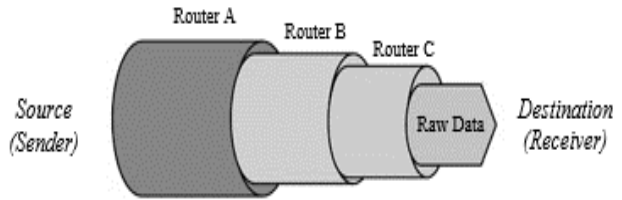

Fig. 1. Onion Routing encapsulates data from source to destination

*B. Garlic Routing*

Garlic Routing [11, 24] is a methodology using by I2P Anonymous Network to hide the identity of sender and receiver of data packets by bundling multiple messages into a layered encryption structure. It is a variation of onion routing that clumps messages together much like a garlic bulb has many cloves. The layered "onion" strategy of Tor implies that a single packet is encrypted more than once but it is still a single message. This strategy of Tor makes timing observations easier that are a method to correlate a Tor entry and exit node. I2P bundles messages together in a packet where each message is like a clove hanging off a garlic bulb. Tor's routing is bi-directional, meaning that data to and from the destination take the same path through Tor. Unlike Tor, garlic routing is uni-directional which data take one path to get to the destination site and a different path to send data back to the requester. This makes observation more troublesome since it's not conceivable to know what path the other half of the discussion is taking. Garlic routing is a variant of onion routing that encrypts multiple messages together to make it more difficult for attackers to perform traffic analysis and to increase the speed of data transfer. Fig. 2 shows the simple structure of Garlic routing where packets are considered as chips with a few pieces of data, where each piece of data is also a garlic with data pieces.

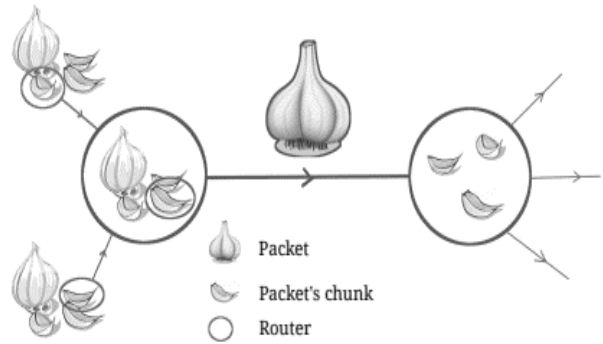

Fig. 2. Garlic Routing of data packets in a network

## C. Sidechains

Sidechains [25] are an essential innovation in the blockchain field with some interesting long-term implications and effects on the broader interoperability and scalability of blockchain networks. Adding new functionalities to the existing blockchain will make this technology realizes its potential more widely. The idea of sidechains originally come to extend the functionality of interoperable blockchain networks, where data can be sent and received between interconnected blockchain networks. Sidechain enables data flow between two blockchain networks in a decentralized manner to transfer and synchronize tokens between two chains. Fig. 3 depicts a core blockchain connecting to a sidechain with a two-way peg protocol [26], which allows direct transfers of a cryptocurrency from the main blockchain to a second blockchain and vice versa.

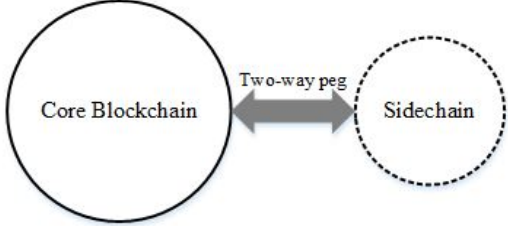

Fig. 3. A sidechain connected to a core blockchain

## IV. PROPOSED SOLUTION

We introduce the idea of PET-Enabled Sidechain (PETES) with the aim to improve the privacy of transactions in blockchain-based applications, where sidechains can operate as a shield on stored data in the core blockchain. By using the knowledge of two-way peg protocol in enabling transactions between two blockchains, we propose a Privacy Assurance Module (PAM) to manage data flow according to predefined privacy policies based on data privacy requirements.

Fig. 4 shows the conceptual view of our proposed PETES structure to ensure privacy policy of data stored in the core blockchain. PAM receives a published transaction and sends it to a PET-enabled sidechain which supports a peer-to-peer customized PET to deanonymize the transaction identity.

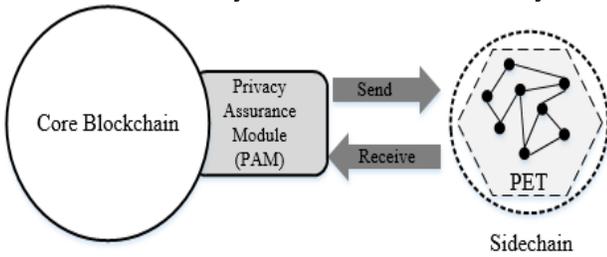

Fig. 4. Proposed PETES structure to ensure privacy policy

According to the figure, Privacy Assurance Module (PAM) sends and receives transactions in two-way peg protocol. A smart contract [27], [28] is a self-execute agent that automatically follows contract rules, and will be resided on blockchain for immutability. PETES consists of many smart contracts which are aware of our proposed Garlic Onion Routing (GOR) technique. A sender does not broadcast the new transaction, PAM receives it and initializes a path for the transaction from Entry node to the Exit node. Note that PAM is in charge of broadcasting the final transaction after doing security process in sidechain. Fig. 5 shows the combination of smart contracts with GOR technique with the purpose of higher data privacy and information preservation. The process of data anonymization of smart contracts will be done with a GOR technique in the sidechain.

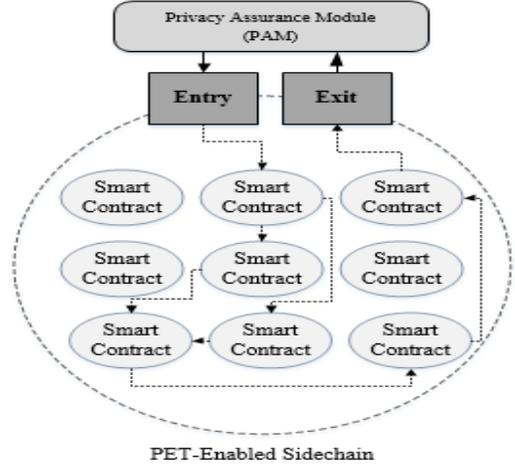

Fig. 5. Inside our PET-Enabled sidechain

Every smart contract on GOR is directly connected (or has the potential to connect) with every other smart contract, meaning that a GOR packet can be relayed from and to potentially any other smart contract. This is not where payment channels do not fully-connect the entire network, and where the network topology is publicly known for routing smart contract. Data fusion of the network topology and the small amount of information from GOR packets may still be enough to uncover information in certain circumstances. For example, if a smart contract has only a single payment channel connection going to one intermediate, then any payments sent to and from the smart contract wallet will have to pass through the intermediate node, which would be able to obtain a lot of information about the wallet node's payments regardless of the GOR used. The degree of confidentiality and privacy provided for transactions that take place on sidechains depends on what technology the sidechain uses, in our case we have used GOR technique. If a sidechain fails or get hacked, it will not damage core blockchain. So damages will be limited within the sidechain. This has allowed people to use sidechains to experiment with pre-release versions of blockchain technologies and sidechains with different permissions to the primary blockchain. Smart contracts will execute based on PAM policies in the sidechain. Also, if we need to make a change in the smart contract, we will be able to apply it to the sidechain. Privacy protocols can be easily integrated at the sidechain level.

The privacy protection mechanism in PETES structure of our proposed GOR is more formally presented in Algorithm 1. In this algorithm, $n$ is the count of the network paths from core blockchain to sidechain that the PAM selects for data transfer. The symbol $dn$ is a different packet's chunk that can be sent in $n$ directions in a simultaneous manner. The symbol $t$ is the time of data transfer inside the involved transactions or smart contracts. $S_K$ is the smart contracts allocated by the sidechain to core blockchain in accordance with PAM, and $C_{Ki}$ is the cryptocurrency key for layers in multiple paths in the GOR technique. With the help of GOR technique, sidechain could prevent fraud and sending inaccurate information on the transaction and smart contracts through managing data flow according to predefined privacy policies in PAM.

Privacy-preservation in PETES allows to search and access specific data blocks while hiding through GOP technique. For example, if A sends a file or asset to user B, both parties can see the transaction details, but C can see that A and B have been transacting, but they cannot see the details of the transferred asset that has an effective role in GOP algorithm provided in PETES.

| **Algorithm 1.** *Garlic Onion Routing (GOR)* |
| --- |
| Sidechain = **Receive** (Transactions & Smart-Contracts (SC)) |
| PAM→ **Send** (Policy) |
| Sidechain→ **Checking** (SC) |
| If (Smart-Contracts == untrust) |
|    { **Modify** (SC) |
|    Core_Blockchain = **two-way peg** ( SC)} |
| Sidechain= **Record** (field_T) // Recording some of features from transaction based on PAM |
| **H_trans_id** (*n*) // hide transaction identity |
|    **Elliptic Curve Cryptography (ECC)** (pkt_chunk) // is adopted to encrypt |
|    Sidechain = **allocate** ($S_K$) |
| Sidechain →**Encrypt_layer** (send data ($d_n$)) // encrypt and send packet's chunks to Core-blockchain |
| $C_{Ki}$= **ECC** (*n, t*) // cryptocurrency key for layers in multiple paths |
|  Sidechain → Send_PKT_chunk// based on PAM policy |
|    Core-blockchain → **Received** (Transactions & SC) |
| Core-blockchain = **Decrypt** (Transactions & SC) // using private key and re-organize data |
| PAM → Updating (Policy) |
|    End |

## V. CONCLUSIONS AND FUTURE WORK

This paper proposes a privacy pervasive sidechain as a new blockchain-specific PET (named PETES) by integrating Garlic Routing and Onion Routing into a Garlic GOR framework suitable to the structure of blockchains. We introduce a PAM to manage data flow according to predefined privacy policies. PETES structure works beside core blockchain, also PETES structure is able to enhance data privacy and modify smart contracts. In the sidechain, we would be able to protect anonymity and confidentiality of users through the GOR with regard to the PAM policy.

For near future work, the implementation and evaluation of the proposed framework is of the highest priority. We plan to implement PETES on a trust computing platform to be able to increase data privacy in open blockchains.